\documentclass{raa}           

\usepackage{graphicx,times}
\usepackage{natbib}
\usepackage{amssymb,amsmath}
\usepackage{amsfonts}
\usepackage[breaklinks,colorlinks,urlcolor=blue,citecolor=blue,linkcolor=blue]{hyperref}

\begin{document}

   \title{Transonic accretion and winds around Pseudo-Kerr black holes and comparison with general relativistic solutions}

 \volnopage{ {\bf 20XX} Vol.\ {\bf X} No. {\bf XX}, 000--000}
   \setcounter{page}{1}

   \author{Abhrajit Bhattacharjee\inst{}, Sandip K. Chakrabarti\inst{} and Dipak Debnath\inst{}
   }

   \institute{Indian Centre for Space Physics, 43 Chalantika, Garia Station Road, Kolkata 700084, India; {\it abhrajitb12@gmail.com, sandipchakrabarti9@gmail.com, dipakcsp@gmail.com}\\
\vs \no
   {\small Received 20XX Month Day; accepted 20XX Month Day}
}

\abstract{Spectral and timing properties of accretion flows on a black hole depend on their density and temperature 
distributions, which in turn come from the underlying dynamics. Thus, an accurate description of the flow which includes hydrodynamics and radiative transfer 
is a must to interpret the observational results. In the case of non-rotating black holes, Pseudo-Newtonian description of surrounding space-time  enables one to make a significant progress in predicting spectral and timing properties. This formalism is lacking for spinning black holes. In this paper,  we show that there exists an exact form of `natural' potential 
derivable from the general relativistic (GR) radial momentum equation. Use of this potential in an otherwise Newtonian set of equations allows to describe transonic flows very accurately  
as is evidenced by comparing with solutions obtained from the full GR framework. We study the properties of the critical points and the centrifugal 
pressure supported shocks in the parameter space spanned by the specific energy and the angular momentum, and compare with the results of GR hydrodynamics. We show that this potential can safely be used for the entire range of Kerr parameter $-1<a<1$ for modeling 
of observational results around spinning black holes. We assume the flow to be inviscid. Thus, it is non-dissipative with constant energy and angular momentum. These assumptions are valid very close to the black hole as the infall timescale is much shorter as compared to the viscous timescale.
\keywords{accretion, accretion disks --- black hole physics --- hydrodynamics --- shock waves --- relativity
}
}

   \authorrunning{A. Bhattacharjee, S. K. Chakrabarti and D. Debnath }            %author_head in even pages
   \titlerunning{Accretion and winds in Pseudo-Kerr geometry}  % title_head in odd pages
   \maketitle

%________________________________________________ sections below
% 
\section{Introduction}

The accretion of matter onto black holes is widely believed to be a fundamental physical process  that has played a significant role in the theories of quasars, active galactic nuclei (AGN) and compact binary systems (e.g., \citealt{lynden1969galactic, pringle1972accretion, pringle1981accretion}). It has been extensively studied during the past five decades. Early studies of an accretion disk employed the standard thin disk model of \citet{shakura1973black} that used the Newtonian potential to describe the gravitational field of the black hole and assumed Keplerian distribution of accreting matter. The standard disk model is incomplete in the sense that the radial drift velocity of the accreting matter is neglected and the inner boundary condition at the horizon was not taken care of, and instead, the inner edge of the disk was chosen to coincide with the marginally stable orbit, i.e., three Schwarzschild radius, for non-rotating black holes. The relativistic version of the standard disk model was formulated by \citet{novikov1973astrophysics} though it still has the same limitations. The entire flow is assumed to be subsonic even though the black hole accretion is necessarily transonic and definitely supersonic on the horizon. Although the standard disk model could explain the thermal component of the black hole accretion disk spectrum, it is unable to explain the non-thermal power-law component. With the observation of high-energy X-rays and $\gamma$-rays it is realised that a typical accretion flow configuration cannot be made of a standard disk alone and it should have a hot component also which could be a rapidly falling, transonic, low viscosity flow in the form of an advective corona of the standard disk.

In regions very close to a black hole, the general relativistic (GR) effects play a crucial role. However, to avoid the complexity of general relativity, the pseudo-Newtonian approach has been devised that allows one to use the equations of Newtonian hydrodynamics but with pseudo-potentials instead of the Newtonian potential that aspire to mimic the corresponding relativistic effects. In the case of accretion around a non-rotating black hole, \citet[][hereafter PW80]{paczynsky1980thick} proposed a pseudo-Newtonian potential that captured the salient features of the space-time around a Schwarzschild black hole quite accurately. In particular, it correctly reproduces the locations of both the marginally stable and marginally bound circular orbits, though the specific energies at these orbits deviated by a few percent. This potential has been used in numerous works to study the structure and physical properties of accretion flows around a non-rotating black hole (\citealt{matsumoto1984viscous, abramowicz1988slim, kato1988pulsational, chakrabarti1989standing, yang1995shock, chakrabarti1995spectral, hawley2001global}) with a good deal of success.

In reality, however, most of the astrophysically relevant black holes are expected to possess considerable angular momentum and they can even be close to maximally spinning with $a=1$, $a$ being the angular momentum per unit mass of the black hole (\citealt{bardeen1970kerr}). Therefore, the important feature that needs to be incorporated in models of accretion disks is the rotation of the black hole. However, solving the hydrodynamic equations which include radial and rotational motions using the full GR framework is itself a challenging task. When the viscous effects, magnetic fields and radiative transfer 
are included, a proper formulation of the problem is often very difficult. In such cases, a sufficiently accurate result with all the salient features with lesser
computational complications become very handy and may motivate one to carry out numerical
simulations to study stabilities of those solutions. Carrying out spectral fits using each data set with theoretical  models  and eventually studying evolution of physical parameters of the flow using massive number of data sets,  requires a formalism which should not be time consuming. This should also be accurate enough so that the errors crept in for not using strict
GR equations are within the observational errors. Several attempts have been made to propose pseudo-Newtonian potentials to study accretion processes around Kerr black holes (e.g., \citealt{chakrabarti1992newtonian, artemova1996modified, semerak1999pseudo, mukhopadhyay2002description, chakrabarti2006studies}). All these potentials were formulated with specific constraints and they have their own limitations. 

In this paper, we derive the \textit{exact} form of the effective potential by identifying a  term  as the radial force in the radial momentum equation in the corotating frame in Kerr geometry. This potential correctly reproduces the locations of both the marginally stable and marginally bound orbits in the full range of the spin parameter and is free from any a priori constraints. The same form of the potential was used earlier by \citet{dihingia2018limitations} for relativistic accretion flows. However, we treat the effective potential as a {\it true} pseudo-Kerr (PK) potential in the sense that except for borrowing this potential from GR, we use 
the same formalism as was used in \citet[hereafter C89]{chakrabarti1989standing} where hybrid model was studied in Schwarzschild geometry. We study the transonic solutions and classify 
the parameter space based on the number of critical points and the type of flow topology. We also discuss the formation of shocks in accretion and in winds and present a comparison of our results with those obtained from the full GR formalism for various sets of parameters. We find that all GR results in Kerr geometry are reproduced with very high accuracy. One successful flow solution which explain the spectral  properties of accretion flows is the two component  (Keplerian and sub-Keplerian) advective flows as presented in \citet[hereafter CT95]{chakrabarti1995spectral}. The so-called quasi-periodic oscillations (QPOs) become a natural consequence of resonance oscillation of the shocks  (\citealt{chakrabarti2015resonance}). A proper fit of data 
requires the rates of the two components as well as the shock location and the compression
ratio of the flow across it. Since the latter quantities are reproduced accurately using our
pseudo-Kerr potential, the procedure of CT95 can be used in Kerr geometry and the spin 
can be determined along with the black hole mass from the fit of each data set.

In the next section, we present the basic equations governing a thin, axisymmetric accretion flow in the equatorial plane of the Kerr geometry. In Section 3, we carry out the critical point analysis and classify the parameter space based on the multiplicity of critical points and the nature of flow solutions. In Section 4, we investigate shock formation in accretion and winds and compare with general relativistic results. Finally, in Section 5, we make the concluding remarks. 

\section{Basic flow equations}

\subsection{The background metric and the model assumptions of the flow}

We consider a thin, stationary and axisymmetric accretion flow in the equatorial plane of a Kerr black hole. We adopt the geometrized units $G=M=c=1$, where $G$ is the gravitational constant, $M$ is the mass of the black hole and $c$ is the speed of light, such that the units of mass, length and time are $M$, $GM/c^2$ and $GM/c^3$ respectively. A polytropic equation of state is chosen for the accreting matter, $p=K(s)\rho^\gamma$, where $p$ and $\rho$ are the isotropic pressure and the matter density respectively, $\gamma$ is the adiabatic index (assumed to be constant throughout the flow, and is related to the polytropic index $n$ by $\gamma=1+1/n$) and $K(s)$ is a measure of  the specific entropy $s$ of the flow which remains constant for an adiabatic flow. The entropy, and thus $K(s)$, can change only across a shock due to generation of entropy.

We choose cylindrical coordinates $(t,r,\phi,z)$ and vertically integrate the flow equations. The half-thickness of the flow, $H(r)$, is assumed to be much smaller than the cylindrical radius, i.e., $H(r)\ll r$. The vacuum metric in and near the equatorial plane of a Kerr black hole is of the form (\citealt{novikov1973astrophysics})
\begin{equation}
    ds^2=-\frac{r^2\Delta}{A}dt^2+\frac{A}{r^2}(d\phi-\omega dt)^2+\frac{r^2}{\Delta}dr^2+dz^2,
\end{equation}
where $a$ is the spin parameter, $\Delta=r^2-2r+a^2$, $A=r^4+r^2a^2+2ra^2$ and $\omega=2ar/A$.

The event horizon of the black hole is located at $r=1+(1-a^2)^{1/2}$, i.e., at the outer root of $\Delta=0$. The self-gravity of the disk matter is neglected and the central plane of the accretion disk is assumed to coincide with the equatorial plane of the black hole.

\subsection{Relativistic hydrodynamics and the effective potential}

We assume a perfect fluid characterized by pressure $p$, rest mass density $\rho$ and specific internal energy $\varepsilon$, defined in the local rest frame of the fluid. We may write the stress-energy tensor of the fluid as
\begin{equation}
    T_{\mu\nu}=h\rho u_\mu u_\nu+pg_{\mu\nu},
\end{equation}
where
\begin{equation}
    h=1+\varepsilon+p/\rho
\end{equation}
is the specific enthalpy. The unit flow vector $u_\mu$ satisfies the normalization condition $u_\mu u^\mu=-1$ which is conserved along the flow lines.

In a stationary $(\partial/\partial t=0)$ and axisymmetric $(\partial/\partial\phi=0)$ space-time the conserved quantities are
\begin{equation}
    \mathcal{E}=h u_t \quad\textrm{and}\quad \mathcal{L}=-h u_\phi,
\end{equation}
where $\mathcal{E}$ and $\mathcal{L}$ are conserved energy and angular momentum respectively, $u_t$ is the specific binding energy and $u_\phi$ is the azimuthal component of the unit flow vector.

The basic equations of relativistic hydrodynamics are the energy-momentum conservation equations 
 \begin{equation}
     \nabla_\mu T^{\mu\nu}=0
 \end{equation}
and the continuity equation
\begin{equation}
    \nabla_\mu(\rho u^\mu)=0.
\end{equation}
By projecting Eq. (5) onto the space orthonormal to the unit flow vector using the projection tensor, $h_{\mu\nu}=g_{\mu\nu}+u_\mu u_\nu$, yields the relativistic Euler equations
\begin{equation}
    h\rho u^\mu\nabla_\mu u^\nu+(g^{\mu\nu}+u^\mu u^\nu)\partial_\mu p=0.
\end{equation}
The radial component of this equation can be put into the form (\citealt{chakrabarti1996global}, hereafter C96)
\begin{equation}
    u\frac{du}{dr}+\frac{1}{r\Delta}\left(a^2-r+\frac{A\gamma_\phi^2B}{r^3}\right)u^2+\frac{A\gamma_\phi^2B}{r^6}+\frac{1}{h\rho}\left(\frac{\Delta}{r^2}+u^2\right)\frac{dp}{dr}=0,
\end{equation}
where 
\begin{equation*}
    u=u^r,
\end{equation*}
\begin{equation*}
    \gamma_\phi=\left[1-\frac{A^2}{\Delta r^4}(\Omega-\omega)^2\right]^{-1/2},
\end{equation*}
and
\begin{equation*}
    B=(\Omega a-1)^2-\Omega^2r^3.
\end{equation*}

Here, $\Omega=u^\phi/u^t$ is the angular velocity of the flow with respect to the stationary observer and $l=\mathcal{L}/\mathcal{E}=-u_\phi/u_t$ is the conserved specific angular momentum of the flow. The quantity $B$ can be expressed as
\begin{equation}
    B=\left(1-\frac{\Omega}{\Omega_{+}}\right)\left(1-\frac{\Omega}{\Omega_{-}}\right)
\end{equation}
in terms of the angular velocities $\Omega_\pm=\pm(r^{3/2}{\pm}a)^{-1}$ corresponding to prograde ($+$) and retrograde ($-$) Keplerian orbits, respectively.

It is convenient to choose a frame that corotates with the same angular velocity $\Omega$ of the flow, the corotating frame, in which the radial velocity is defined by (\citealt{abramowicz1996advection})
\begin{equation}
    \gamma_v v=\frac{v}{(1-v^2)^{1/2}}=u^rg_{rr}^{1/2},
\end{equation}
so that the radial velocity in the corotating frame is given by
\begin{equation}
    v=\left(1+\frac{\Delta}{r^2u^ru^r}\right)^{-1/2}.
\end{equation}
This implies $v=1$ at the horizon $(\Delta=0)$, i.e., the infall velocity at the horizon equals the speed of light, independent of the mass and spin of the black hole. Since the sound speed is always much less than the speed of light, the flow is supersonic at the horizon. This velocity 
can be defined in other ways (e.g., $V$ in  C96). Thus, \textit{black hole accretion process is necessarily transonic in nature} (see also \citealt{chakrabarti1990theory}).

In the corotating frame, Eq. (8) takes the form
\begin{equation}
    \gamma_v^2v\frac{dv}{dr}+\frac{A\gamma_\phi^2B}{r^4\Delta}+\frac{1}{h\rho}\frac{dp}{dr}=0.
\end{equation}
In analogy with Newtonian hydrodynamics, the second term can be identified as the gradient of the effective potential at the equatorial plane of Kerr geometry, such that the radial force is given by
\begin{equation}
    F(r)=-\frac{d\Phi_\mathrm{eff}}{dr}=-\frac{A\gamma_\phi^2B}{r^4\Delta}.
\end{equation}
The corresponding effective potential obtained from the force is
\begin{equation}
    \Phi_\mathrm{eff}(r)=1+\frac{1}{2}\ln\left(\frac{r(r^2-2r+a^2)}{r^3+(a^2-l^2)r+2(a-l)^2}\right).
\end{equation}
This potential correctly reproduces the locations of both the marginally stable and marginally bound orbits for the entire range of the spin parameter in Kerr geometry. The corresponding
gravitational potential $\Phi$ obtained by using $l=0$ in the above expression of $\Phi_\mathrm{eff}$ is used in the calculation of the vertical height of the flow. Some other potential obtained in more complex way (see e.g., \citealt{dihingia2018limitations}) also agrees with the one given above which is obtained in a straightforward manner when proper frame is chosen.

\subsection{Equations governing a transonic flow}

We assume a steady-state and axisymmetric accretion flow in equilibrium in a direction transverse to the flow at the equatorial plane of a Kerr black hole. We do not consider any  dissipation in the flow, so the specific angular momentum $l$ of the accreting matter remains constant. In the non-relativistic limit, the accretion flow is characterized by $v\ll1$ (i.e., the Lorentz factor $\gamma_v=1$) and specific enthalpy $h\sim1$ all throughout. Thus the radial momentum equation in the steady state in the pseudo-Kerr formalism becomes,
\begin{equation}
v\frac{dv}{dr} + \frac{1}{\rho}\frac{dp}{dr} + \frac{d\Phi_\mathrm{eff}}{dr}=0 .
\end{equation}
This is of the same form as in C89 written for the Schwarzschild geometry, except that the
final term now contains the gravity, centrifugal, coriolis and spin-orbit coupling 
terms in a non-linear way. The conserved specific energy of the flow is obtained by integrating the radial momentum equation: 
\begin{equation}
    \mathcal{E}=\frac{v^2}{2}+\frac{a_s^2}{\gamma-1}+\Phi_\mathrm{eff},
\end{equation}
where, $a_s$ is the sound speed and $\Phi_\mathrm{eff}$ is the effective potential (Eq. 14). Since we are dealing with a thin flow in vertical equilibrium, the vertical component of flow velocity is very small as compared to the radial component. The continuity equation can be integrated to obtain, apart from a geometric factor, the mass conservation equation given by (C89)
\begin{equation}
    \dot{M}={\rho}vrH(r),
\end{equation}
where, 
\begin{equation}
    H(r)=a_sr^{1/2}(\Phi')^{-1/2}
\end{equation}
is the half-thickness of the flow that is obtained by equating the vertical component of the force due to the gravitational potential $\Phi$ (i.e., $\Phi_\mathrm{eff}$ for $l=0$) and the pressure gradient force. Here, prime ($'$) represents the derivative $d/dr$ with respect to $r$ in flat geometry. It is useful to express the mass conservation equation (Eq. 17) in terms of the flow velocity and the sound speed as (C89)
\begin{equation}
    \dot{\mathcal{M}}=a_s^{2n+1}vr^{3/2}(\Phi')^{-1/2}.
\end{equation}

This quantity $\dot{\mathcal{M}}\propto K^n\dot{M}$ is the entropy accretion rate of the flow that changes only at the shocks due to generation of entropy (C89). The Rankine-Hugoniot shock conditions presented for strictly one dimensional flow (\citealt{landau1959fluid}) are modified in the following way (C89):

\noindent the continuity of energy flux across the shock
\begin{equation}
    \mathcal{E}_{+}=\mathcal{E}_{-},
\end{equation}
the continuity of mass flux across the shock
\begin{equation}
    \dot{M}_{+}=\dot{M}_{-},
\end{equation}
and the pressure balance condition
\begin{equation}
    W_{+}+\Sigma_{+}v_{+}^2=W_{-}+\Sigma_{-}v_{-}^2.
\end{equation}
Here, the subscripts ``$-$" and ``$+$" indicate, respectively, the quantities just before and after the shock, whereas, $W$ and $\Sigma$ are the pressure and the matter density, integrated in the vertical direction (see e.g., \citealt{matsumoto1984viscous}) given by,
\begin{equation}
    \Sigma=\int_{-H}^{H} \rho\mathrm{d}z=2\rho I_{n}H
\end{equation}
and
\begin{equation}
    W=\int_{-H}^{H} p\mathrm{d}z=2p I_{n+1}H,
\end{equation}
where, $I_n=(2^{n}n!)^2/(2n+1)!$ and $n$ is the polytropic index. The modification (C89) of Rankine-Hugoniot relations was needed to ensure that the weakest shock in a flow with vertical equilibrium has unit compression ratio.

\section{Critical point analysis}

 The radial velocity gradient of the accretion flow obtained by differentiating Eqs. (16) and (19) with respect to $r$ and eliminating $da_s/dr$ is given by,
 \begin{equation}
     \frac{dv}{dr}=\left(\frac{na_s^2(3\Phi'-r\Phi'')}{(2n+1)r\Phi'}-\Phi_\mathrm{eff}'\right)\bigg/\left(v-\frac{2na_s^2}{(2n+1)v}\right).
 \end{equation}

Since the flow is subsonic ($v<a_s$) at large radii and crosses the horizon supersonically ($v>a_s$) into the black hole, the denominator ($\mathcal{D}$) of Eq. (25)  must vanish at an intermediate location. Therefore, to have a smooth solution across that location, the numerator ($\mathcal{N}$) must also vanish to keep $dv/dr$ finite. Such a location is the critical point ($r_c$) or sonic point of the flow. Thus, $dv/dr$ is well-defined and regular at the critical point only if $\mathcal{N}=\mathcal{D}=0$ and this yields the following critical point conditions:
\begin{equation}
    v_c^2=\left(\frac{2n}{2n+1}\right)a_{sc}^2
\end{equation}
and
\begin{equation}
    na_{sc}^2=\frac{(2n+1)r_c\Phi_\mathrm{eff}'\Phi'}{3\Phi'-r_c\Phi''},
\end{equation}
where the subscript `c' denotes the quantities evaluated at the critical points. Note that  since $a_{sc}^2$ is positive, we must have $\Phi_\mathrm{eff}' > 0$ at the critical point. This implies that $B>0$ in Eq. (12). This puts a constraint on the angular momentum of the flow and from Eq. (9), one finds that this condition boils down to the fact that the flow has to be sub-Keplerian at the critical points, i.e., $l<l_\mathrm{Kep}$. In all the cases of inviscid flows studied in \citet{chakrabarti1990theory}, this was shown in the context of accretion around non-rotating black holes using the PW80 potential.

From Eq. (26), one can calculate the Mach number at the critical point as,
\begin{equation}
    M_c=\frac{v_c}{a_{sc}}=\sqrt{\frac{2n}{2n+1}}.
\end{equation}
This is of the same form as the flow in vertical equilibrium in full general relativity (C96) and also using PW80 potential for flows around non-rotating black holes (C89).

Since the radial velocity gradient has the form, $dv/dr=0/0$ at $r=r_c$, we apply the l'Hospital rule to calculate $(dv/dr)_c$ at the critical point as
\begin{equation}
    \left(\frac{dv}{dr}\right)_c=\frac{-B\pm\sqrt{B^2-4AC}}{2A},
\end{equation}
where,
\begin{subequations}
\begin{align}
    A &= \frac{4(n+1)}{2n+1},\\
    B &= \frac{4}{2n+1}\sqrt{G\Phi_\mathrm{eff}'},\\
    C &= \frac{2}{2n+1}G\Phi_\mathrm{eff}'-\frac{G'}{G}\Phi_\mathrm{eff}'+\Phi_\mathrm{eff}'',
\end{align}
\end{subequations}
where $G=(3\Phi'-r_c\Phi'')/2r_c\Phi'$. Using the critical point conditions in Eq. (16), one can calculate the conserved specific energy $\mathcal{E}$ of the flow as a function of the critical point location $r_c$ and specific angular momentum $l$ of the flow. Depending on the choice of the parameters ($\mathcal{E}, l, a$), the accretion flow may possess multiple critical points through which the matter enters into the black hole. The nature of the critical points depends on the radial velocity gradients at the critical points. The critical point is saddle-type when the values of $(dv/dr)_c$ are real and of opposite signs, whereas the critical point is center-type when $(dv/dr)_c$ becomes imaginary.  

For a flow with angular momentum above a critical value, i.e., $l>l_c$, the flow possesses three critical points, two saddle-type critical points flanking a center-type critical point. The origin of the center-type critical point is attributed to the angular momentum of the flow, whereas the existence of the inner saddle-type critical point is a purely general relativistic effect (\citealt{liang1980transonic}). It is necessary for a globally acceptable transonic solution, although not sufficient, that the flow must pass through at least one saddle-type critical point. However, for a significant region of the parameter space the flow can also experience a discontinuity or a shock transition after passing through the outer critical point first. The post-shock flow subsequently passes through the inner critical point before entering the horizon of the black hole.

\begin{figure}
    \centering
    \includegraphics[scale=0.55]{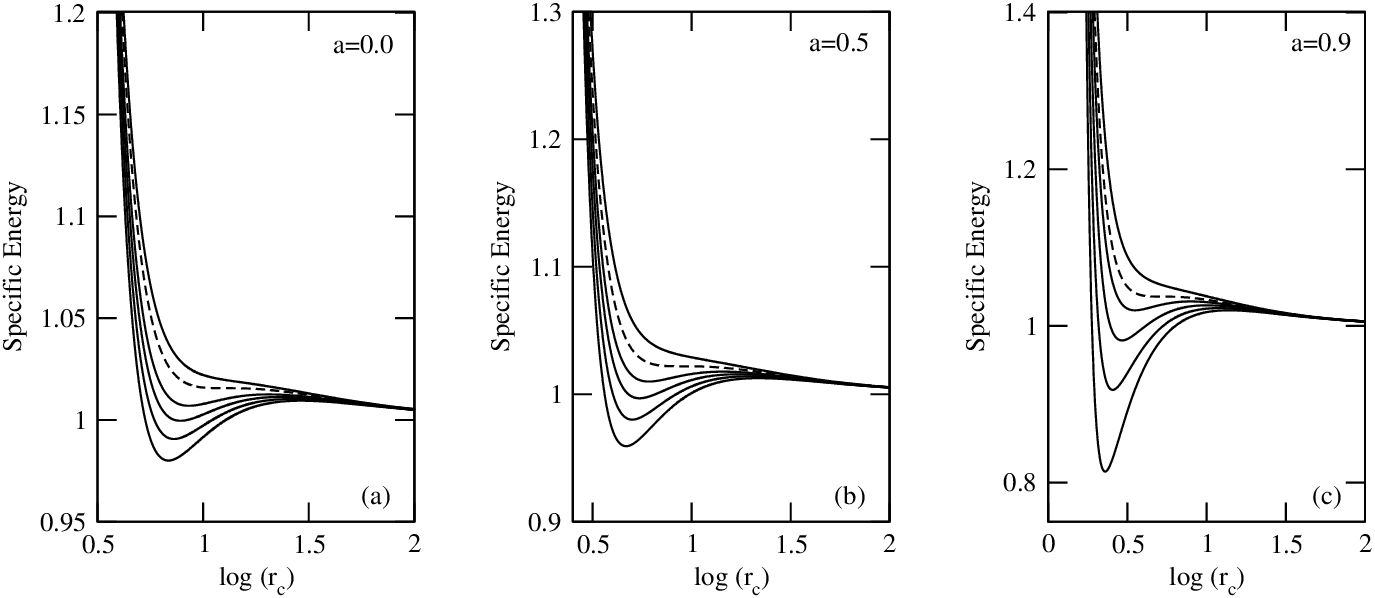}
    \caption{Variation of the specific energy of the flow with the location of critical points for (a) $a=0.0$, (b) $a=0.5$ and (c) $a=0.9$ for different values of specific angular momentum using our PK potential. The curves are drawn for (starting from the topmost curve), (a) $l=2.8, 2.935, 3.1, 3.2, 3.3, 3.4$, (b) $l=2.4, 2.567, 2.7, 2.8, 2.9, 3.0$ and (c) $l=2.0, 2.121, 2.2, 2.3, 2.4, 2.5$, respectively. Dashed lines are drawn for critical angular momenta $l_c$ such that for $l<l_c$, there is only one critical point in the flow.}
    \label{fig:Erplot}
\end{figure}

In Figs. 1a-1c, we plot the variation of the specific energy of the flow as a function of the location of the critical points for various angular momenta for the spin parameters (a) $a=0.0$, (b) $a=0.5$ and (c) $a=0.9$. The specific angular momenta for which the curves have been drawn are, from the topmost curve, (a) $l=2.8, 2.935, 3.1, 3.2, 3.3, 3.4$, (b) $l=2.4, 2.567, 2.7, 2.8, 2.9, 3.0$ and (c) $l=2.0, 2.121, 2.2, 2.3, 2.4, 2.5$ respectively. The dashed curves in all the cases correspond to the critical angular momentum $l_c$ such that for $l<l_c$ the number of critical points is always one, whereas for $l>l_c$ three critical points are possible depending on the specific energy of the flow. For a particular value of the specific energy of the flow and spin parameter of the black hole, the critical points are obtained as the intersections of the constant energy (horizontal) lines with the constant angular momentum curves. Critical points with negative slope of the curve correspond to saddle-type critical points, whereas positive slope correspond to the center-type critical points.

\begin{figure}
    \centering
    \includegraphics[scale=0.55]{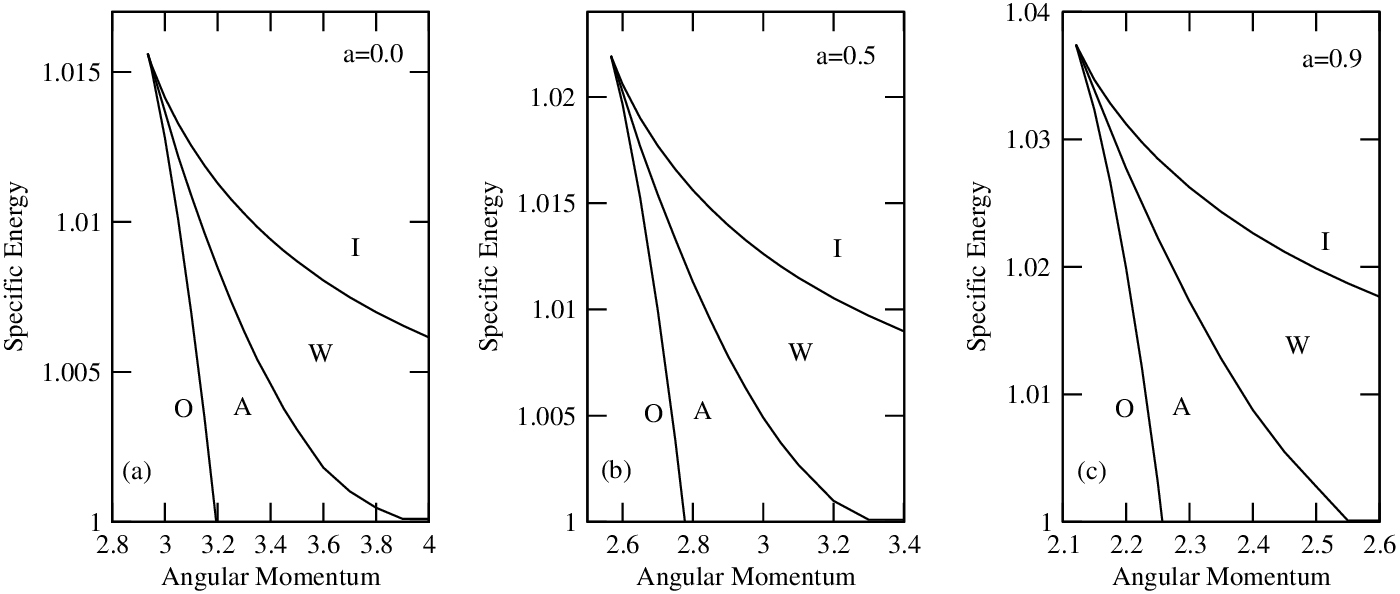}
    \caption{Division of the parameter space according to the different types of flow solutions using our PK potential for (a) $a=0.0$, (b) $a=0.5$ and (c) $a=0.9$. The regions are marked as O, A, W, and I depending on the multiplicity of the critical points. Flows with parameters from the region O and I have only one critical point, whereas the regions A and W correspond  to accretion and wind solutions that allow multiple critical points.}
    \label{fig:Elplot}
\end{figure}

\begin{figure}
    \centering
    \includegraphics[scale=0.55]{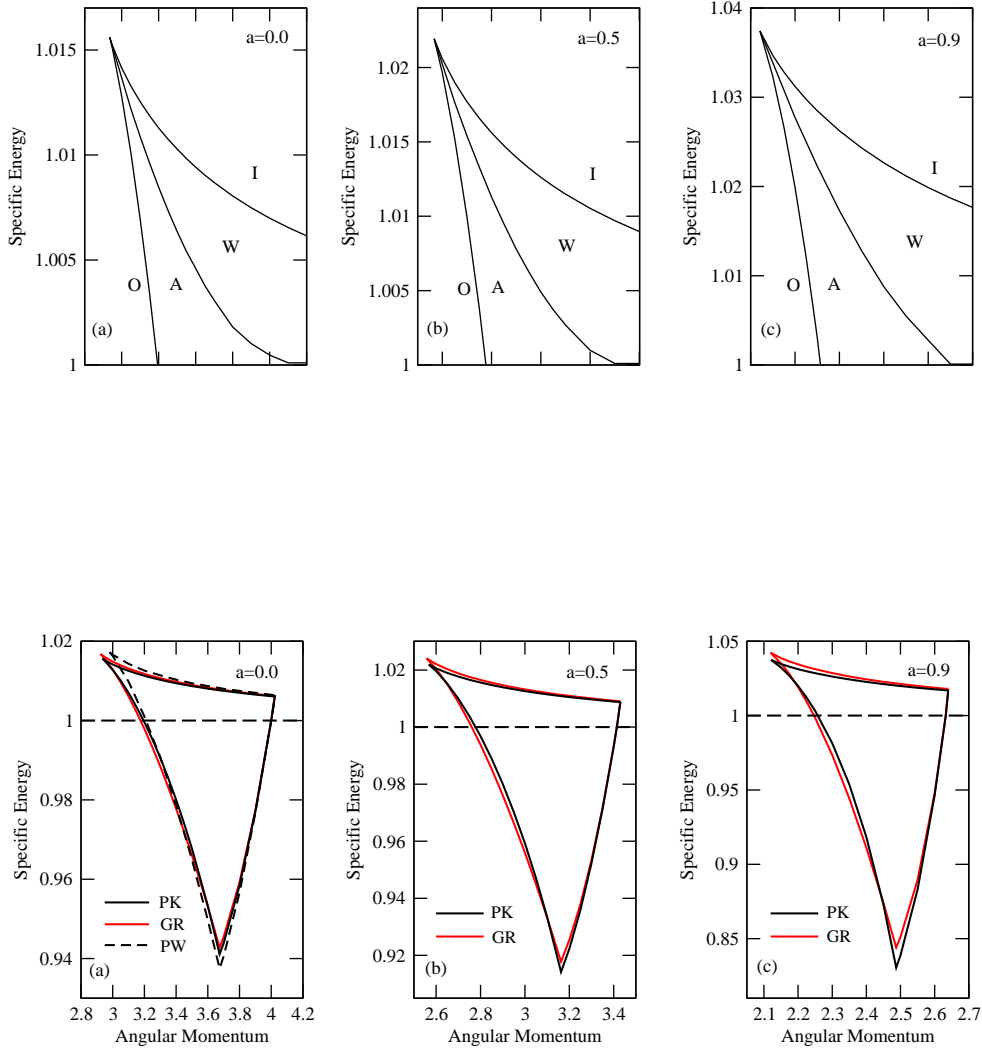}
    \caption{Comparison of the entire parameter space for multiple critical points using our PK potential with full GR approach (C96) for (a) $a=0.0$, (b) $a=0.5$ and (c) $a=0.9$. For comparison, in (a), we also plotted the parameter space obtained using the PW potential (PW80).}
    \label{fig:Elfull}
\end{figure}

In Figs. 2a-2c, we classify the parameter space spanned by ($\mathcal{E},l$) for the flow solutions that allow multiple critical points for the spin parameters (a) $a=0.0$, (b) $a=0.5$ and (c) $a=0.9$. Four distinct regions of the parameter space are identified, marked as O, A, W and I, based on the number of critical points and the type of the solution topology. The solution with parameters from the regions O and I has only the outer and the inner critical points respectively. On the other hand, parameters from the regions A and W correspond to the solutions that contain two saddle-type critical points and a center-type critical point in between them. In general, the entropy of the two saddle-type critical points are different, except on the curve separating the regions A and W where the entropy is the same. In the region A, the inner critical point has a greater entropy than the outer critical point and shocks may form only for accretion flows, whereas in the region W, the outer critical point has a greater entropy than the inner critical point and shocks may form only for winds. In Figs. 3a-3c, we present the comparison of the entire parameter space for multiple critical points using our pseudo-Kerr approach with the parameter space obtained using full GR for the spin parameters (a) $a=0.0$, (b) $a=0.5$ and (c) $a=0.9$. For $a=0.0$, we also compare the parameter space obtained using the PW80 potential. We find that the parameter space using the pseudo-Kerr approach is in good agreement with that obtained from GR for the entire range of the spin parameter.

\begin{figure}
    \centering
    \includegraphics[scale=0.4]{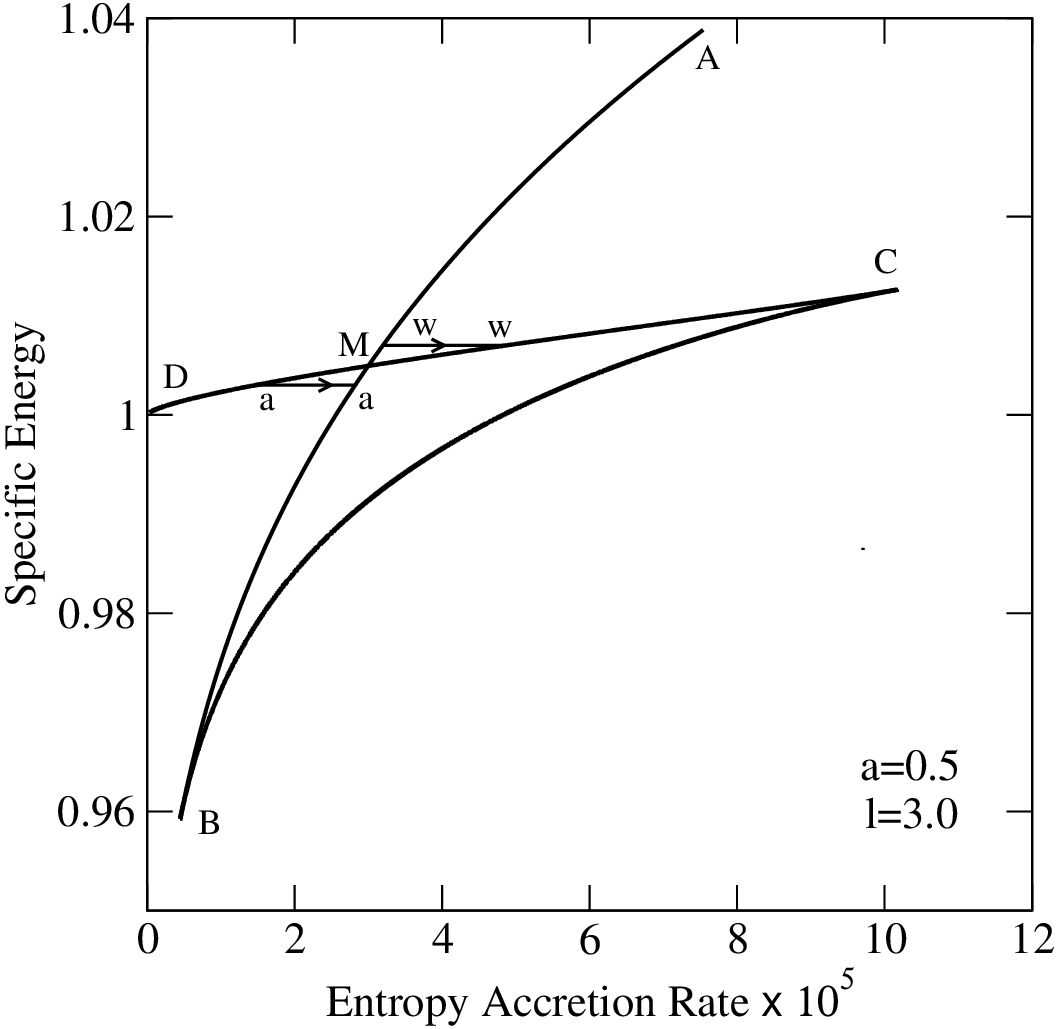}
    \caption{Plot of the variation of the specific energy of the flow as a function of the entropy accretion rate for $a=0.5,l=3.0$. Flow with parameters from the segment AMB passes through the inner saddle-type critical point, whereas the flows with parameters from the segment CMD passes through the outer saddle-type critical point. The segment BC corresponds to the unphysical, center-type critical point.}
    \label{fig:EMdot}
\end{figure}

\section{Solutions containing shocks}

To classify the shock solutions based on the flow parameters, it is convenient to obtain a quantity that remains invariant at the shock locations (C89, C96). In order to calculate the shock invariant, we rewrite the energy conservation equation (Eq. 20), the entropy accretion rate equation (Eq. 19) and the pressure balance equation (Eq. 22) in terms of the Mach number $M=v/a_s$ of the flow as
\begin{equation}
    \frac{1}{2}M_+^2a_{s+}^2 + na_{s+}^2 = \frac{1}{2}M_-^2a_{s-}^2 + na_{s-}^2,
\end{equation}
\begin{equation}
    \mathcal{\dot{M}}_+ = \frac{M_+a_{s+}^{2(n+1)}r_s^{3/2}}{\sqrt{\Phi'(r_s)}},
\end{equation}
\begin{equation}
    \mathcal{\dot{M}}_- = \frac{M_-a_{s-}^{2(n+1)}r_s^{3/2}}{\sqrt{\Phi'(r_s)}},
\end{equation}
and
\begin{equation}
    \frac{a_{s+}^\nu}{\mathcal{\dot{M}}_+}\left(\frac{2\gamma}{3\gamma-1}+\gamma M_+^2\right) = \frac{a_{s-}^\nu}{\mathcal{\dot{M}}_-}\left(\frac{2\gamma}{3\gamma-1}+\gamma M_-^2\right),
\end{equation}
where, $\nu=(3\gamma-1)/(\gamma-1)$ and $r_s$ is the shock location. From Eqs. (31)-(34), it becomes clear that only the entropy is discontinuous across a shock, and we obtain the shock invariant quantity as 
\begin{equation}
    C = \frac{[M_+(3\gamma-1)+(2/M_+)]^2}{2+(\gamma-1)M_+^2} = \frac{[M_-(3\gamma-1)+(2/M_-)]^2}{2+(\gamma-1)M_-^2}.
\end{equation}

In Fig. 4, we plot the variation of the specific energy $\mathcal{E}$ with the entropy accretion rate $\mathcal{\dot{M}}$ at all the critical points of the flow for $a=0.5$ and $l=3.0$. The branches AMB and CMD correspond to the inner and the outer saddle-type critical points, respectively, and the branch BC corresponds to the center-type critical point. However, a flow can pass smoothly through both the saddle-type critical points simultaneously only if the flow makes a discontinuous transition in between them. It is important to note that the flow generally prefers to pass through a shock because the entropy of the subsonic branch of the post-shock flow is higher as compared to the entropy of the supersonic branch of the pre-shock flow. Shocks in accretion flows are possible only when the entropy at the inner critical point is greater than that of the outer critical point, i.e., when the condition $\mathcal{\dot{M}}_\mathrm{in}>\mathcal{\dot{M}}_\mathrm{out}$ is satisfied. For shocks in winds, the corresponding condition is $\mathcal{\dot{M}}_\mathrm{in}<\mathcal{\dot{M}}_\mathrm{out}$. Thus, from the Figure, it is clear that shocks in accretion are possible only for the horizontal transitions below M, where the curves AB and CD intersect, and shocks in winds are possible only for the horizontal transitions above M. Thus, as a consequence of the second law of thermodynamics, a shock always connect two flow solutions with different entropy and the post-shock entropy must be higher than the pre-shock entropy. Examples of shock transitions for two different values of the specific energy are also shown in the Figure as the horizontal arrows $aa$ and $ww$ for two different values of the specific energy.

\begin{figure}
    \centering
    \includegraphics[scale=0.6]{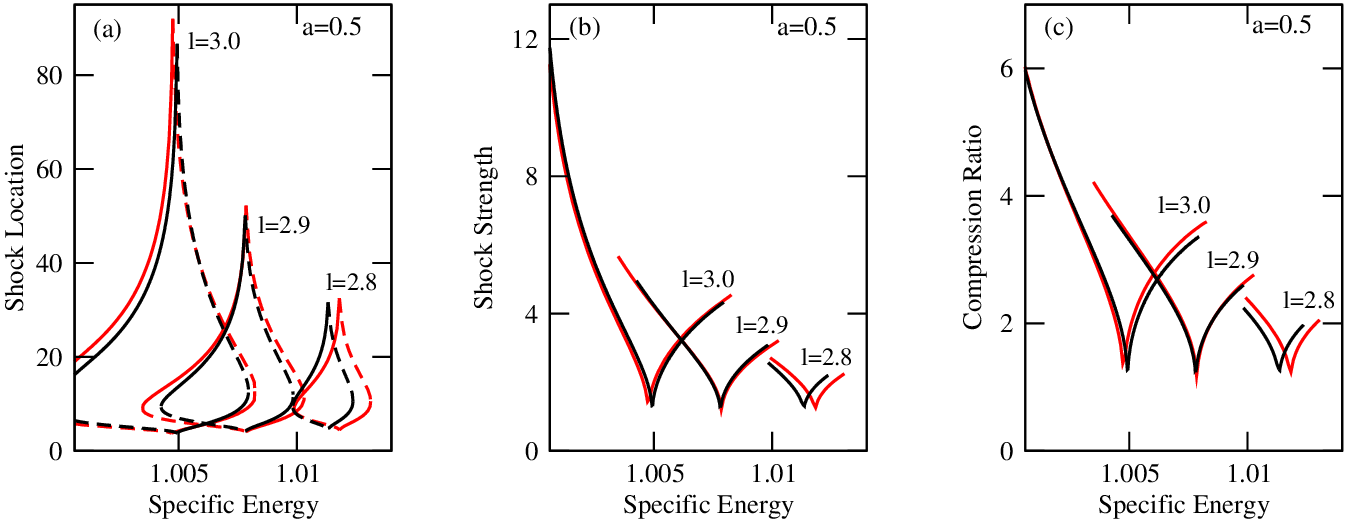}
    \caption{Variation of (a) shock location, (b) shock strength and (c) compression ratio with the specific energy of the flow in accretion and winds for different angular momenta $l=2.8,2.9,3.0$. The spin parameter is chosen to be $a=0.5$. The stable shock locations are plotted as solid curves and the dashed curves represent the unstable shock locations. The left and the right segment of each set of curves for a particular angular momentum correspond to accretion and winds, respectively. We also compare the results obtained using full GR (C96), shown in red color.}
    \label{fig:rsE}
\end{figure}

In C89 (see also \citealt{chakrabarti1990theory}), it was found that in inviscid flows,
the shock conditions are satisfied at four different locations, denoted by $r_\mathrm{s1},r_\mathrm{s2},r_\mathrm{s3}$ and $r_\mathrm{s4}$. However, only $r_\mathrm{s2}$ and $r_\mathrm{s3}$ were found to be useful for the study of shocks in accretion and winds. It was also shown that only $r_\mathrm{s3}$ is stable for accretion, whereas only $r_\mathrm{s2}$ is stable for winds (\citealt{chakrabarti1993smoothed}). At the shock $r=r_\mathrm{s}$, the shock strength ($\mathcal{S}$) is the ratio of the pre- and post-shock Mach numbers, and the compression ratio ($\mathcal{R}$) is the ratio of the post- and pre-shock vertically integrated matter density. In Figs. 5a-5c, we show the variation of (a) shock location, (b) shock strength and (c) compression ratio for $a=0.5$, as a function of the specific energy of the flow. Each set of curves are plotted for different values of the specific angular momenta (marked on the set), where the left and right segments of each set corresponds to accretion and winds, respectively. Further, the stable shock locations are represented by solid curves and the unstable shock locations by dashed curves. A comparison with the GR values are also presented, shown in red color. It may be noted that the stable shocks in accretion are formed at larger radii when the specific energy of the flow is increased. One can also observe that with increase in angular momentum the centrifugal barrier becomes stronger and stable shocks are located farther away from the black hole.

\begin{figure}
    \centering
    \includegraphics[scale=0.55]{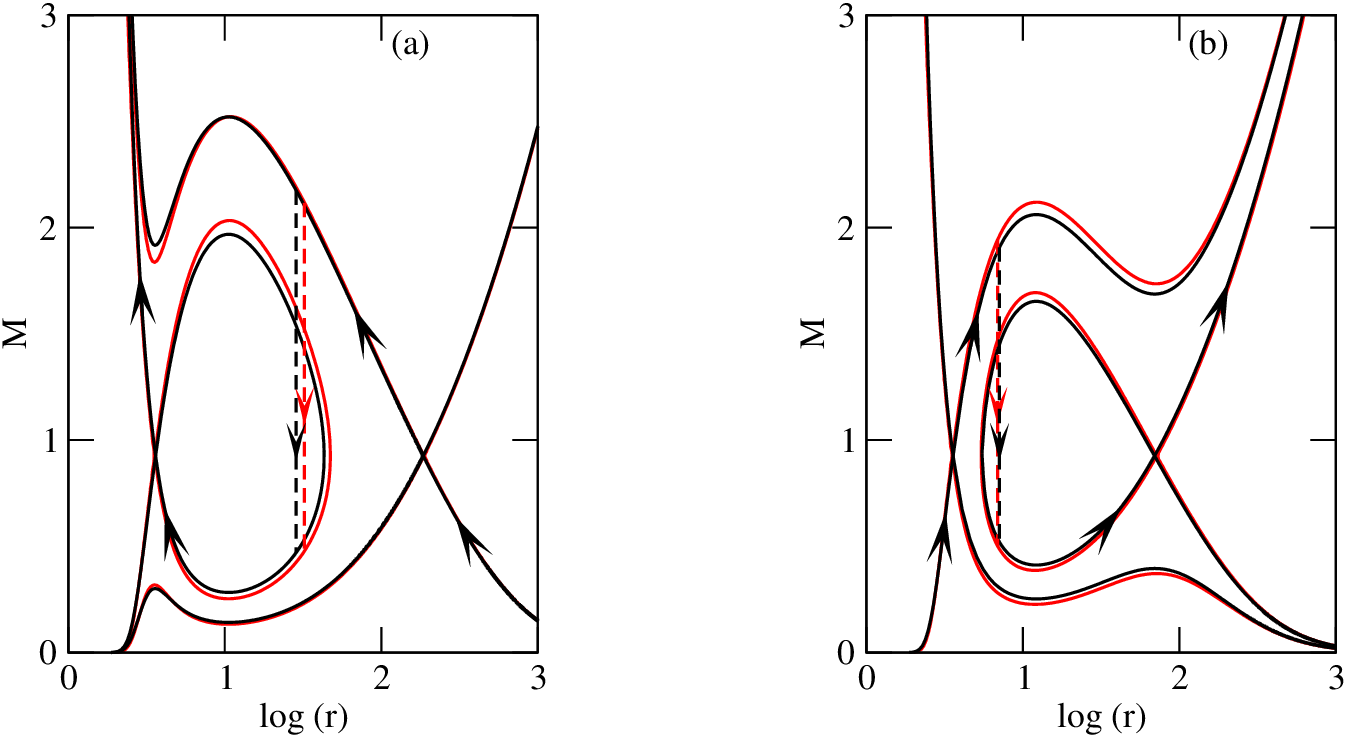}
    \caption{Comparison of the shock solutions in (a) accretion and (b) winds for the parameters (a) $a=0.5,l=3.0,\mathcal{E}=1.003$ and (b) $a=0.5,l=3.0,\mathcal{E}=1.007$ with full GR solutions (C96), shown in red color. The vertical arrows indicate the stable shock transitions $r_\mathrm{s3}$ for accretion and $r_\mathrm{s2}$ for winds, respectively.}
    \label{fig:shock0p5}
\end{figure}

We now present shock solutions for accretion and winds for two sets of parameters and compare with the full GR solutions (C96). We illustrate shock solutions where the Mach number ($M=v/a_s$) of the flow is plotted as a function of the radial distance. In Fig. 6a, we choose the parameters $a=0.5$, $l=3.0$ and $\mathcal{E}=1.003$. The two saddle-type critical points and the corresponding values of the entropy accretion rates are found to be $r_\mathrm{in}=3.61$, $r_\mathrm{out}=185.30$, $\mathcal{\dot{M}}_\mathrm{in}=2.819\times 10^{-5}$ and $\mathcal{\dot{M}}_\mathrm{out}=1.475\times 10^{-5}$. Since $\mathcal{\dot{M}}_\mathrm{in}>\mathcal{\dot{M}}_\mathrm{out}$, a stable shock is formed in accretion and the shock location is found to be $r_\mathrm{s3}=28.58$ of shock strength $\mathcal{S}=4.63$. The corresponding GR values are $r_\mathrm{in}=3.50$, $r_\mathrm{out}=186.81$, $\mathcal{\dot{M}}_\mathrm{in}=2.74\times 10^{-5}$, $\mathcal{\dot{M}}_\mathrm{out}=1.491\times 10^{-5}$, $r_\mathrm{s3}=32.29$ and $\mathcal{S}=4.43$. In Fig. 6b, we choose the parameters $a=0.5$, $l=3.0$ and $\mathcal{E}=1.007$. The two saddle-type critical points and the corresponding values of the entropy accretion rates are found to be $r_\mathrm{in}=3.58$, $r_\mathrm{out}=69.44$, $\mathcal{\dot{M}}_\mathrm{in}=3.197\times 10^{-5}$ and $\mathcal{\dot{M}}_\mathrm{out}=4.868\times 10^{-5}$. Since $\mathcal{\dot{M}}_\mathrm{in}<\mathcal{\dot{M}}_\mathrm{out}$, a stable shock is formed in winds and the shock location is found to be $r_\mathrm{s2}=7.28$ of shock strength $\mathcal{S}=3.66$. The corresponding GR values are $r_\mathrm{in}=3.47$, $r_\mathrm{out}=71.11$, $\mathcal{\dot{M}}_\mathrm{in}=3.12\times 10^{-5}$, $\mathcal{\dot{M}}_\mathrm{out}=5.001\times 10^{-5}$, $r_\mathrm{s2}=6.89$ and $\mathcal{S}=3.88$.  As a consequence of the second law of thermodynamics which requires that the post-shock entropy must be higher than the pre-shock entropy, i.e., $\mathcal{\dot{M}}_{+}>\mathcal{\dot{M}}_{-}$, the parameters for stable shocks in accretion and winds are mutually exclusive (C89, C96). Note that although the flow has a shock-free solution (passing through $r_\mathrm{out}$ for accretion and through $r_\mathrm{in}$ for winds), the flow chooses to pass through a shock because the shock solution is of higher entropy.

\begin{figure}
    \centering
    \includegraphics[scale=0.55]{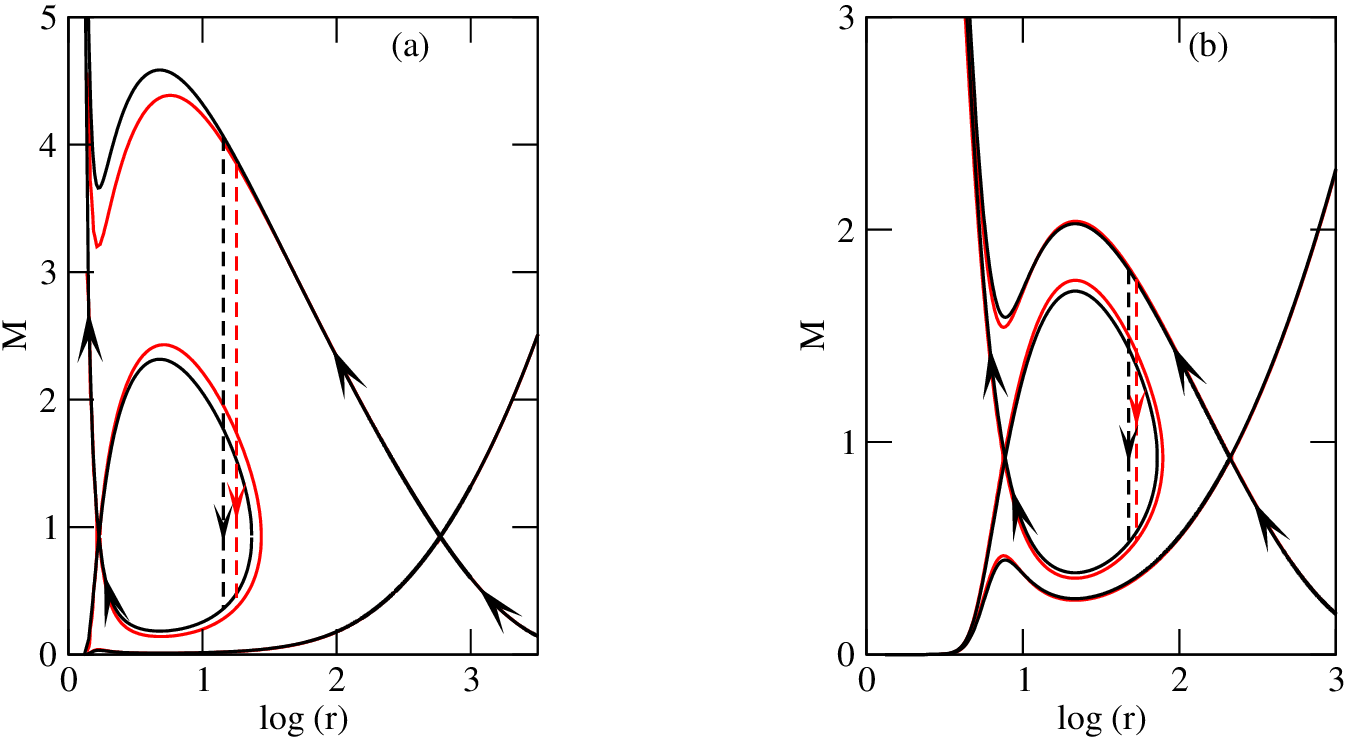}
    \caption{Comparison of accretion shock solutions in (a) prograde flow and (b) retrograde flow for the parameters (a) $a=0.95,l=2.3,\mathcal{E}=1.001$ and (b) $a=-0.95,l=4.0,\mathcal{E}=1.0025$ with full GR solutions (C96), shown in red color. The stable shock transitions $r_\mathrm{s3}$ for accretion flows are indicated by the vertical arrows in both the cases.}
    \label{fig:shock0p95}
\end{figure}

We now study the nature of shock solutions in prograde and retrograde accretion flows, and compare with the full GR solutions (C96). In Fig. 7a, we show a shock solution in prograde flow with parameters $a=0.95$, $l=2.3$ and $\mathcal{E}=1.001$. The two saddle-type critical points and the corresponding values of the entropy accretion rates are found to be $r_\mathrm{in}=1.69$, $r_\mathrm{out}=592.84$, $\mathcal{\dot{M}}_\mathrm{in}=4.784\times 10^{-5}$, $\mathcal{\dot{M}}_\mathrm{out}=2.962\times 10^{-6}$ and the stable shock is located at $r_\mathrm{s3}=14.32$ of shock strength $\mathcal{S}=11.31$. The corresponding GR values are $r_\mathrm{in}=1.66$, $r_\mathrm{out}=594.32$, $\mathcal{\dot{M}}_\mathrm{in}=4.108\times 10^{-5}$, $\mathcal{\dot{M}}_\mathrm{out}=2.973\times 10^{-6}$, $r_\mathrm{s3}=18.29$ and $\mathcal{S}=10.44$. In Fig. 7b, we show a shock solution in retrograde flow with parameters $a=-0.95$, $l=4.0$ and $\mathcal{E}=1.0025$. The two saddle-type critical points and the corresponding values of the entropy accretion rates are found to be $r_\mathrm{in}=7.72$, $r_\mathrm{out}=210.42$, $\mathcal{\dot{M}}_\mathrm{in}=1.509\times 10^{-5}$, $\mathcal{\dot{M}}_\mathrm{out}=1.086\times 10^{-5}$ and the stable shock is located at $r_\mathrm{s3}=47.47$ of shock strength $\mathcal{S}=3.44$. The corresponding GR values are $r_\mathrm{in}=7.43$, $r_\mathrm{out}=211.98$, $\mathcal{\dot{M}}_\mathrm{in}=1.478\times 10^{-5}$, $\mathcal{\dot{M}}_\mathrm{out}=1.097\times 10^{-5}$, $r_\mathrm{s3}=53.41$ and $\mathcal{S}=3.30$. The shock locations for a retrograde flow are found to be farther away from the black hole than that for a prograde flow. Thus, we find that the results obtained using the pseudo-Kerr approach is in close agreement with the corresponding GR results.

We observe that the parameter space for multiple critical points gradually shifts towards the higher specific energy and lower angular momentum edge as the spin parameter is increased. This also means that the energy range for allowed shock transition increases with the spin parameter. Clearly this occurs due to spin-orbit coupling and this reveals the role of the centrifugal barrier in the formation of shocks. Thus, the spin of the black hole plays a crucial role in determining the structure of accretion disks.

\section{Concluding remarks}

Explanation of the iron-line features and the temporal properties of accretion flows around black holes 
often require that the black holes have a significant amount of intrinsic spin, e.g., GRO J1655-40 (\citealt{reis2009determining}), V404 Cygni (\citealt{walton2017living}) and EXO 1846-031 (\citealt{draghis2020new}).  One way to obtain the spins would be to identify the
quasi-periodic oscillation (QPO) frequencies in softer spectral states when the accretion flow is dominated by the Keplerian disk (see \citealt{reynolds2021observational}, for a recent review). In the Two Component 
Advective Flow (TCAF) paradigm of CT95, spectral and timing properties of 
massive and supermassive black holes are explained using hydrodynamics and radiative transfer
around Schwarzschild black holes where PW80 potential is used. Since 
the boundary of the Compton cloud is always found to be far away from the black holes in harder states,
fitting of parameters remained accurate even when the spin parameter is significant.  However,
to fit the observational data as the spectral state becomes softer, one requires that the equations are 
solved very close to a Kerr black hole. This is a very challenging task in full GR formalism. Thus a pseudo-Newtonian
approach for a spinning black hole would be very useful, especially when the evolution of flow parameters 
are studied through fitting a large quantity of observed data.

While a large number of such potentials are present in the literature, and some efforts have been made
to use them in obtaining transonic flow properties in Kerr geometry (e.g., \citealt{mondal2006studies, dihingia2018limitations}),  we found that these were either not applicable for all spins, 
or were used incorrectly, mixing relativistic and non-relativistic equations, defeating the spirit
of a pseudo-potential approach. The latter cases are
difficult to assess since it is uncertain as to what extent results are influenced by the potential or the relativistic equations. 
Also, the potentials were not derived in a straight forward way and mostly put in an `ad hoc' manner. 
In the present paper, we derived the potential from the radial force term in the general 
relativistic radial momentum equation written in the corotating frame.  
This was done so as to have the
closest possible analogy with the respective Newtonian force components. The potential was obtained 
directly integrating the force term while `Newtonize' the other terms of the same equation using the 
limit of unit Lorentz factor.
Thus the novel aspect of our work is that we used a truly Newtonian approach, which includes usage of an appropriate potential in non-relativistic Newtonian fluid dynamic equations and yet we
could reproduce the transonic properties of a flow in vertical equilibrium as obtained in the context of full GR (C96) very accurately. In particular, classification of the parameter space based on the number of critical points of the flow, the nature of the solution topologies and calculated shock locations are all with a few percentage of error even when we used the near extreme Kerr parameter. The shock strengths and the compression ratios were also found to be 
very similar. It is to be noted that the shock location directly gives the size of the Compton 
cloud and along with the shock strength and accretion rates, one gets the optical 
depth and temperature of the cloud (CT95), so essential to obtain the power-law spectrum. These
flow properties, which are the basis of the TCAF paradigm, 
remain very similar to those obtained in full GR formalism. 
We found that the shocks in accretion flows around high spin black holes will 
form much closer to the black hole as compared to those for non-rotating black holes, and therefore the QPOs would have higher frequencies (\citealt{chakrabarti2015resonance}). Also, the angular 
momentum required 
to form shocks for rapidly spinning black holes is much lower due to the spin-orbit coupling. Our result for a non-rotating black hole is, in fact, better than those obtained using PW80 potential. Therefore, the effective potential $\Phi_\mathrm{eff}$ can be used to study accretion flow around a Kerr black hole for the entire range of the spin parameter $-1<a<1$. Thus our formalism paves the way to fit data with the TCAF solution and extract both mass and spin parameters of the black hole in the system.

In this paper, we have studied inviscid accretion flows in the equatorial plane of Kerr geometry using the pseudo-Kerr formalism presented here. Although, in reality, accretion is driven by the viscosity due to turbulent motion of the matter in the disk, the infall timescale is very short compared to the viscous timescale (the characteristic timescale for angular momentum transport) in a region close to the black hole where the space-time curvature plays a significant role. This means that there is practically no dissipation of the angular momentum during the radial drift of matter in such a region (\citealt{kozlowski1978analytic}) and the flow behaves like an inviscid flow in some sense. Away from the black hole, the viscosity plays a major role and the transport of angular momentum is possible within the infall timescale. In fact, as in earlier results with the PW80 potential (\citealt{chakrabarti1996grand}), we anticipate that in the presence of viscosity, the flow will have weaker shocks and the central critical points will be of `spiral'- or `nodal'- type as opposed to `circle'-type. In \citet{chakrabarti1996grand}, it was shown that with high enough viscosity, the flow becomes a Keplerian accretion disk while at the same time passing through the innermost saddle-type critical point. It is of interest to check whether this behavior still persists in Kerr geometry. In a future work, we will address these issues using our pseudo-Kerr formalism. The results would be reported elsewhere.

% Authors can give a citation as `\citealt{Michel+etal+1992}'.
% You may also use \cite, \citep and \citet for citation, and use Table~1
% or Figure~1 and so forth. Using \ref and \label for cross-references of
% Tables/Figures is a good way in adjusting/adding/removing text, tables or
% figures.

\normalem
\begin{acknowledgements}
AB acknowledges a grant of the ISRO sponsored RESPOND project (ISRO/RES/2/418/18-19).
Research of SKC and DD is supported in part by the Higher Education Dept. of the Govt. of West Bengal, India. The authors would like to thank the anonymous referee for valuable comments and suggestions.
\end{acknowledgements}
  
\bibliographystyle{raa}

\end{document}